\documentclass[12pt]{article}
\usepackage{amsmath}
\usepackage{graphicx,psfrag,epsf}
\usepackage{enumerate}
\usepackage{natbib}
\usepackage{url} 

\usepackage{graphicx}
\usepackage{slashbox}
\usepackage{amssymb}
\usepackage[english]{babel}
\usepackage{amssymb,amsmath,multirow,array}
\usepackage[english]{babel}

\newcommand{\blind}{0}

\date{}
\begin{document}


\def\spacingset#1{\renewcommand{\baselinestretch}%
{#1}\small\normalsize} \spacingset{1}


  \title{Parametric bootstrapping in a generalized extreme value regression model for binary response}
  \author{
	\textbf{DIOP Aba}\footnote{Corresponding author}\\
  \textit{E-mail: aba.diop@uadb.edu.sn} \\
  \textit{Equipe de Recherche en Statistique et Modèles Aléatoires} \\
  \textit{Université Alioune Diop, Bambey, Sénégal} \\
  \textbf{DEME El Hadji} \\
  \textit{E-mail: elhadjidemeufrsat@gmail.com} \\
  \textit{Laboratoire d'Etude et de Recherche en Statistique et Développement} \\
  \textit{Université Gaston Berger, Saint-Louis, Sénégal}  
  }
  
\maketitle
\if0\blind

\bigskip
\begin{abstract}
Generalized extreme value (GEV) regression is often more adapted when we investigate a relationship between a binary response variable $Y$ which represents a rare event and potentiel predictors $\mathbf{X}$. In particular, we use the quantile function of the GEV distribution as link function. Bootstrapping assigns measures of accuracy (bias, variance, confidence intervals, prediction error, test of hypothesis) to sample estimates. This technique allows estimation of the sampling distribution of almost any statistic using random sampling methods. Bootstrapping estimates the properties of an estimator by measuring those properties when sampling from an approximating distribution. In this paper, we fitted the generalized extreme value regression model, then we performed parametric bootstrap method for testing hupthesis, estimating confidence interval of parameters for generalized extreme value regression model and a real data application.
\end{abstract}

\noindent%

\spacingset{1.45} 
\section{Introduction}
\label{sec:intro}
Classical methods of statistical inference do not provide correct answers to all the concrete problems the user may have. They are only valid under some specific application conditions (e.g. normal distribution of populations, independence of samples etc.). The estimation of some characteristics such as dispersion measures (variance, standard deviation), confidence intervals, decision tables for hypothesis tests, is also based on the mathematical expressions of the probability laws, as well as approximations of these when the calculation was not feasible.

A good estimate of the nature of the population distribution can lead to powerful results. However, the price to pay is high if the assumption of the distribution is incorrect. It is therefore important to consider other analysis methods (such as non-parametric methods, for which the conditions of application are more restrictive), which are more flexible with the choice of the distribution and based on this, bootstrap methods was introduced.

Bootstrapping is a relatively new, computer-intensive statistical methodology introduced by \cite{efron79}. The bootstrap method replaces complex analytical procedures by computer intensive empirical analysis. It relies heavily on Monte Carlo Method where several random resamples are drawn from a given original sample. The bootstrap method has been shown to be an effective technique in situations where it is necessary to determine the sampling distribution of (usually) a complex statistic with an unknown probability distribution using these data in a single sample. The bootstrap method has been applied effectively in a variety of situations. \cite{efron94}, \cite{shao96} and \cite{shao10} provide a comprehensive discussion of the bootstrap method. \cite{andro20} discussed statistical properties of the approximations of the "bootstrap-generated" data sets in more detail. Many studies have shown that the bootstrap resampling technique provides a more accurate estimate of a parameter than the analysis of any one of the samples (see forexample \cite{carp00}, \cite{zou04}, \cite{man97}, \cite{shao95}).

The bootstrap method is a powerfull method to assess statistical accuracy or to estimate distribution from sample's statistics (\cite{rey04}, \cite{dav03}, \cite{cher99}). In principle there are three different ways of obtaining and evaluating bootstrap estimates: non-parametric bootstrap which does not assume any distribution of the population; semi-parametric bootstrap, which partly has an assumption on the distribution on parameter and whose residuals have no distributional assumption; and finally parametric bootstrap which assumes a particular distribution for the sample at hand. \cite{adj16} have considered parametric and non-parametric bootstrap in the case of classical logistic regression model.

In this work we aim to estimate the probability of infection as function of potential predictors $\mathbf X= \mathbf x$ using a generalized extreme value regression model for binary data, construct confidence intervals and testing hypothesis for the unknown parameters of the modele using both classical and bootstrap (non-parametric and parametric) methods.

The rest of this paper is organized as follows. In Section \ref{sec:method}, we describe the problem of GEV regression model and parametric bootstrapping method. Section \ref{sec:application} present the obtained results. A discussion and some perspectives are given in Section \ref{sec:discussion}.
\section{Method}
\label{sec:method}
\subsection{Generalized extreme value regression model}
Generalized is used to model rare and extreme event (see  \cite{cole01}. In the case where the dependent variable $Y$ represents a rare event, the logistic regression model (obviously used for this category of data) shows relevant drawbacks. We suggest the quantile function of the GEV distribution as link function to investigate the relationship between the binary response variable $Y$ and the potential predictors $\mathbf X$ (see \cite{wang10} and \cite{co13} for more details). We use Bootstrapping method as a tool to estimate parameters and standard errors for generalized extreme value regression model. For a binary response variable $Y_i$ and the vector of explanatory variables $x_i$, let $ \pi(x_i) = \mathbb P(Y_i=1|\mathbf X_i=x_i)$ the conditional probability of infection. Since we consider the class of Generalized Linear Models, we suggest the GEV cumulative distribution function proposed by \cite{co13} as the response curve given by
\begin{equation}\label{eq1sec2}
\begin{split}
\pi(x_i)&= 1-\exp\lbrace\left[(1-\tau(\beta_1+\beta_2 x_{i2}+ \cdots+\beta_p x_{ip}))_{+} \right]^{-1/\tau}  \rbrace \\
&= 1- GEV(-x'_{i}\beta;\tau)
\end{split}
\end{equation}
where $\beta=(\beta_1, \ldots, \beta_p)' \in\mathbb R^p$ is an unknown regression parameter measuring the association between potential predictors and the risk of infection and $GEV(x;\tau)$ represents the cumulative probability at $x$ for the GEV distribution with a location parameter $\mu=0$, a scale parameter $\sigma=1$, an unknown shape parameter $\tau$.\\
For $\tau \rightarrow 0$, the previous model (\ref{eq1sec2}) becomes the response curve of the log-log model, for $\tau > 0$ and $\tau < 0$ it becomes the Frechet and Weibull response curve respectively, a particular case of the GEV one.\\
The link function of the GEV model is given by
\begin{equation}\label{eq2sec2}
\frac{1-\left[\log(1-\pi(x_i)) \right]^{-\tau} }{\tau}=x'_{i}\beta=\eta(x_i)
\end{equation}
The unknown vector parameter $\beta$ will be estimated with $(1-\alpha)$\% confidence intervals $(\alpha \in [0;1])$ and a test of hypothesis $H_0$: $\beta_j = 0$ by both classical approch of GEV regression model and bootstrap methods.
\subsection{Parametric Bootstrapping}
Parametric bootstraps resample a known distribution function, whose parameters are estimated from your sample. A parametric model is fitted using parameters estimated from the distribution of the bootstrap estimates, from which confidence limits are obtained analytically. In applications where the standard asymptotic theory does not hold, the null reference distribution can be obtained through parametric bootstrapping (see \cite{rey04}). Maximum likelihood estimators are commonly used for parametric bootstrapping despite the fact that this criterion is nearly always based upon their large sample behaviour.
\subsubsection{Parametric bootstrap confidence interval}
Using an algorithm by \cite{zou04} or \cite{carp00} a parametric bootstrap confidence interval is obtained as follows:
\vspace{.4cm}

\fbox{\parbox{\textwidth}{
\textit{
\begin{description}
\item[1.] Draw $B$ bootstrap samples $\{(Y_{i}^{(b)},\mathbf X_{i}^{(b)}), i=1,\ldots,n\}$ $(b=1,\ldots,B)$ from the original data sample, and for each bootstrap sample, estimate $\beta$ and $\pi_i$ by its maximum likehood estimators $\widehat \beta_{n}^{(b)}$ in the model \eqref{eq1sec2}.
\item[2.] Calculate the bootstrap mean and standard error of $\widehat \beta_n$ as follows:
\begin{equation*}
\widehat \beta_{n}^{*} = \frac{1}{B}\sum_{b=1}^{B}\widehat \beta_{n}^{(b)} ~~ \text{and} ~~ \widehat s_n = \sqrt{\frac{1}{B-1}\sum_{b=1}^{B}\left( \widehat \beta_{n}^{(b)} - \widehat \beta_{n}^{*}\right)^2 }
\end{equation*}
\item[3.] Calculate $(1-\alpha)100\%$ bootstrap confidence interval by finding quantile of boostrap replicates
\begin{equation*}
\left[ \widehat \beta_{n,L},\widehat \beta_{n,U}\right] = \left[\widehat \beta_{n}^{(b),1-\alpha} ,\widehat \beta_{n}^{(b),\alpha}\right] 
\end{equation*}
\end{description}
}}} 
\vspace{.4cm}

\subsubsection{Parametric Bootstrap for Test of Hypothesis}
We use here the algorithm for parametric test of hypothesis given by \cite{fox15} defined as follows:
\vspace{.4cm}

\fbox{\parbox{\textwidth}{
\textit{
\begin{description}
\item[1.] Estimates parameters $\widehat \beta_{n,j}$ of model \ref{eq1sec2} using the observed data and calculate observed statistic test $t_{\beta_j}^{obs}=\widehat \beta_{n,j} / s_{\widehat \beta_{n,j}}$. Let $\hat{\theta}=(\widehat \beta_{n,j},t_{\beta_j}^{obs})$
\item[2.] Draw $B$ bootstrap samples $\{(Y_{i}^{(b)},\mathbf X_{i}^{(b)}), i=1,\ldots,n\}$ $(b=1,\ldots,B)$ from the original data sample, and for each bootstrap sample, estimate $\beta$ and $t_{\beta_j}^{obs}$ by its maximum likehood estimators $\widehat \beta_{n}^{(b)}$ and $t_{\widehat \beta_{n}^{(b)}}^{obs}$ in the model \eqref{eq1sec2}.
\item[3.] Calculate bootstrap p-value by
\begin{equation*}
p-value=\frac{\sharp\left\lbrace |t_{\widehat \beta_{n}^{(b)}}^{obs}|>|t_{\widehat \beta_{n}}^{obs}| \right\rbrace }{B}
\end{equation*}
\end{description}
}}} 
\vspace{.4cm}
\section{Real data application}
\label{sec:application}
In this setion, we consider a study of dengue fever, which is a mosquitoborne viral human disease.  We consider
here a database of size $n = 515$ (with 15.5\% of 1's), which was constituted with individuals recruited in Cambodia, Vietnam, French Guiana, and Brazil (\cite{dussart11}). Each individual $i$ was diagnosed for dengue infection and coded as $Y_i = 1$ if infection was present and $0$ otherwise. We aim at estimating: i) the risk of infection for those individuals, based on this data set which also includes the following covariate: \textit{Weight} (continuous bounded covariates), ii) testing the impact of weight on infection status by running a generalized extreme value regression analysis of the model defined as follows:
\begin{equation*}
\pi_i = \mathbb{P}(Y_i =1 | \text{\textbf{\textit{Weight}}}) = 1- GEV\left[ -(\beta_1 +\beta_2\times \text{\textbf{\textit{Weight}}});\tau\right] .
\end{equation*}.
\begin{table}[h!]
\tabcolsep=4pt
\def\arraystretch{1.05}
\caption{Parameter estimates of GEV regression model.}
\label{tab1}
%
\begin{center}
\begin{tabular}{ccccccccc}
\hline
Parameter && Estimate && SE && 95\% C.I &&  P-value
\\\hline
Intercept ($\beta_1$) && 0.9947 && 0.0739 && [0.8731,1.1162] &&    \\
Weight ($\beta_2$) && -0.0456 && 0.0012 && [-0.0475,-0.0436] && $< 0.0001$ \\
\hline
\end{tabular}
\parbox{12.1cm}{ ~\\
\underline{Note}: \noindent SE: standard error. C.I: confidence interval}
\end{center}
\end{table}
\begin{table}[h!]
\tabcolsep=4pt
\def\arraystretch{1.05}
\caption{Confidence intervals and p-value by parametric bootstrap.}
\label{tab2}
%
\begin{center}
\begin{tabular}{ccccccccc}
\hline
Parameter && Estimate && SE && 95\% C.I &&  P-value
\\\hline
Intercept ($\beta_1$) && 1.3873 && 0.0926 && [-1.0480,6.7853] &&    \\
Weight ($\beta_2$) && -0.0522 && 0.0015 && [-0.1567,-0.0005] && $< 0.0001$ \\
\hline
\end{tabular}
\parbox{12.1cm}{ ~\\
\underline{Note}: \noindent SE: standard error. C.I: confidence interval}
\end{center}
\end{table}
Results obtained from the GEV regression model \ref{eq1sec2} by parametric bootstrap are shown in Table \ref{tab2}. This results lead to similar conclusion from classical method. It can be observed that the parameter estimates are very close. The standard errors of estimates for parametric bootstrap were slightly bigger compared to that of classical approch. It is observed that in both situations of classical approch and parametric bootstrap method, the effect of weight is highly significant.
\section{Discussion and perspectives}
\label{sec:discussion}
Confidence intervals are good indicators of practical significance, unlike p-values and they also provide more information than p values. Unfortunately, confidence intervals are rarely reported in academic papers. This is because computing confidence intervals are not practical and not possible for some statistics. This is why bootstraps methods, which are resampling techniques for assessing uncertainty, have become popular.\\
In this study we have performed bootstrapping parametric method on a GEV regression model. Moreover bootstrapped method was compared with classical approch while calculating the parameters of GEV regression model. The bootstrap technique used for estimation and testing produced flexible results. Several question can be asked: what is the appropriate value of $B$ for confidence intervalles and for test of hypothesis ? For example \cite{efron94} suggest that $B$ should be between 1000 and 2000 for 90-95 percent confidence intervals. Non-parametric bootstrap method can also be used for this study. With the help of statistical software today it is easy to compute confidence intervals and test of hypothesis for almost any statistics of interest.

\end{document}